\documentclass[12pt]{article}
\usepackage{epsfig,amsfonts,amsthm}
\usepackage[normalem]{ulem}
\usepackage{amsmath,amssymb}
\usepackage{array}
\usepackage{amsmath}
\usepackage{amsfonts}
\usepackage{amssymb}
\usepackage{subfig}
\usepackage{wrapfig}
\usepackage{graphicx}
\usepackage{slashed}
\usepackage{soul}
\newcommand{\be}{\begin{equation}}
\newcommand{\ee}{\end{equation}}
\newcommand{\bea}{\begin{eqnarray}}
\newcommand{\eea}{\end{eqnarray}}

\renewcommand{\Re}{\mathrm{Re }}
\renewcommand{\Im}{\mathrm{Im }}
\newcommand{\doublet}[2]{ \left( \begin{array}{c}#1 \\ #2 \end{array}\right) }

\usepackage{color}

\usepackage[normalem]{ulem}
   
\definecolor{grey}{cmyk}{0,0,0,0.75}
\definecolor{tangerine}{cmyk}{0,0.5,1,0}
\definecolor{darkgreen}{cmyk}{1,0,1,0.23} 
\definecolor{Red}{rgb}{1,0,0}
\definecolor{Blue}{rgb}{0,0,1}
\definecolor{Green}{rgb}{0,1,0}
\definecolor{Grey}{cmyk}{0,0,0,0.75}
\definecolor{Tangerine}{cmyk}{0,0.5,1,0}
\definecolor{Darkgreen}{cmyk}{1,0,1,0.23}
\definecolor{Cyan}{cmyk}{1,0,0,0}
\definecolor{Yellow}{cmyk}{0,0,1,0}

\def\lsim{\mathrel{\rlap{\lower4pt\hbox{\hskip1pt$\sim$}}
    \raise1pt\hbox{$<$}}}         %less than or approx. symbol
\def\gsim{\mathrel{\rlap{\lower4pt\hbox{\hskip1pt$\sim$}}
    \raise1pt\hbox{$>$}}}         %greater than or approx. symbol

\def\beq{\begin{equation}}
\def\eeq{\end{equation}}
\def\bea{\begin{eqnarray}}
\def\eea{\end{eqnarray}}

\def\<{\left\langle}
\def\>{\right\rangle}

\usepackage[normalem]{ulem}

\def\lsim{\mathrel{\rlap{\lower4pt\hbox{\hskip1pt$\sim$}}
    \raise1pt\hbox{$<$}}}         %less than or approx. symbol
\def\gsim{\mathrel{\rlap{\lower4pt\hbox{\hskip1pt$\sim$}}
    \raise1pt\hbox{$>$}}}         %greater than or approx. symbol

\def\beq{\begin{equation}}
\def\eeq{\end{equation}}
\def\bea{\begin{eqnarray}}
\def\eea{\end{eqnarray}}

\def\<{\left\langle}
\def\>{\right\rangle}

\newcommand{\gev}{\mathrm{\;GeV}} 
\newcommand{\bt}{\begin{tabular}}
\newcommand{\et}{\end{tabular}}

\usepackage{hyperref}    
    
\usepackage{tikz}
\usetikzlibrary{decorations.pathmorphing,decorations.markings}

\usepackage{graphicx}%Täl saa laitettuu kuvii. Esim. ps filei. Kun haluu laittaa .epsi tai .eps tiedostoja, niin laittaa eka \includegraphics{nimi.eps}. Sit kääntää pdflatex:lla ja kuva tulee. 

\tikzset{
photon/.style={decorate, decoration={snake,amplitude=2pt, segment length=5pt}, draw=black},
particle/.style={draw=black, postaction={decorate}, decoration={markings,mark=at position .5 with {\arrow[draw=black]{>}}}},
antiparticle/.style={draw=black, postaction={decorate}, decoration={markings,mark=at position .5 with {\arrowreversed[draw=black]{>}}}},
gluon/.style={decorate, draw=black, decoration={coil,amplitude=4pt, segment length=5pt}},
goldstone/.style={draw=green,postaction={decorate},decoration={markings,mark=at position .5 with {\arrow[draw=blue]{>}}}}
}

\begin{document}

\bibstyle{plain}
\bibliographystyle{plain}
\bibdata{plain}

\title{\hfill ~\\[-30mm]
                  \textbf{Lepton collider indirect signatures of dark CP-violation
                   }     }
\date{}

\author{\\[-5mm]
A. Cordero-Cid\footnote{E-mail: {\tt adriana.cordero@correo.buap.mx}}$^{~1}$,\ 
J. Hern\'andez-S\'anchez\footnote{E-mail: {\tt jaime.hernandez@correo.buap.mx}}$^{~1}$,\ 
V.  Keus\footnote{E-mail: {\tt Venus.Keus@helsinki.fi}}$^{~2,3}$,\\ 
%S. ~F.~King\footnote{E-mail: {\tt King@soton.ac.uk}} $^{~3}$,\ \\
S.  Moretti\footnote{E-mail: {\tt S.Moretti@soton.ac.uk}} $^{3,4}$,\
D.  Rojas\footnote{E-mail: {\tt drojas@ifuap.buap.mx}} $^{1,3}$,\  
D.  Soko\l{}owska\footnote{E-mail: {\tt dsokolowska@iip.ufrn.br}} $^{5,6}$
\\ \\
\emph{\small $^1$ Instituto de F\'isica and Facultad de Ciencias de la Electr\'onica,}\\  
\emph{\small 
Benem\'erita Universidad Aut\'onoma de Puebla,
Apdo. Postal 542, C.P. 72570 Puebla, M\'exico,}\\
  \emph{\small $^2$ Department of Physics and Helsinki Institute of Physics,}\\
 \emph{\small Gustaf Hallstromin katu 2, FIN-00014 University of Helsinki, Finland}\\
  \emph{\small $^3$ School of Physics and Astronomy, University of Southampton,}\\
  \emph{\small Southampton, SO17 1BJ, United Kingdom}\\
  \emph{\small  $^4$ Particle Physics Department, Rutherford Appleton Laboratory,}\\
 \emph{\small Chilton, Didcot, Oxon OX11 0QX, United Kingdom}\\
  \emph{\small  $^5$ University of Warsaw, Faculty of Physics, Pasteura 5, 02-093 Warsaw, Poland.}\\
  \emph{\small  $^6$ International Institute of Physics, Universidade Federal do Rio Grande do Norte,}\\
\emph{\small Campus Universitario, Lagoa Nova, Natal-RN 59078-970, Brazil}\\[4mm]
  }

\maketitle

\vspace*{-7mm}

\begin{abstract}
\noindent
{We study an extension of the Standard Model (SM) in which two copies of the SM Higgs doublet are added to the scalar sector. These extra doublets do not develop a vacuum expectation value, hence, they are \textit{inert}. This essentially leads to a 3-Higgs Doublet Model (3HDM) with 2 inert and 1 active scalar doublets, which we denote as  I(2+1)HDM.
We allow for CP-violation in the \textit{inert} sector, where the lightest \textit{inert} state
is protected from decaying to SM particles through the conservation of a $Z_2$ symmetry, so that it is a Dark Matter (DM) candidate. For this scenario, 
we identify  a smoking gun signature of  dark CP-violation in the form of production thresholds of pairs of  \textit{inert} neutral Higgs bosons
 at an $e^+e^-$ collider.\\[15mm]
\begin{center}
{\sl\footnotesize We dedicate this work to the memory of Prof. W. James Stirling, an example to never forget.}
\end{center}
} 
 \end{abstract}

\vspace*{4mm}

\begin{flushright}
IIPDM-2018~~~~~~~~
\end{flushright}

\thispagestyle{empty}
\vfill
\newpage
\setcounter{page}{1}

\section{Introduction}

The Standard Model (SM) of fundamental interactions has been extensively tested in recent decades and the search for its last missing piece -- the SM Higgs particle -- ended in 2012 with  the discovery of a scalar boson with a mass of approximately 125 GeV by ATLAS and CMS experiments at the CERN Large Hadron Collider (LHC) \cite{Aad:2012tfa,Chatrchyan:2012ufa}. Since then, further effort has been spared to study Higgs boson dynamics  at the LHC.  Although the properties of the observed scalar are in agreement with those of the SM Higgs boson, it is still possible that it is just one member of an extended (pseudo)scalar sector. 

There are various reasons why it is generally believed that the SM of particle physics is incomplete. One of the issues that needs to be addressed is the absence of a Dark Matter (DM) candidate in the SM. Cosmological observations imply that about 85\% of matter in the Universe is cold (i.e., non-relativistic at the onset of galaxy formation), non-baryonic,  neutral and weakly interacting \cite{Aghanim:2018eyx}: such a state does not exist in the SM. Various candidates have been proposed so far, the best studied being a Weakly Interacting Massive Particle (WIMP) \cite{Jungman:1995df,Bertone:2004pz,Bergstrom:2000pn}. The mass of this hypothetical particle can vary between a few GeV and a few TeV, however, its exact nature is still unknown.

A particle with such characteristics can come from an extended scalar sector with a discrete symmetry. A well-known example  is the Inert Doublet Model (IDM), a 2-Higgs Doublet Model (2HDM) with an unbroken discrete $Z_2$ symmetry \cite{Deshpande:1977rw}.  The model involves {1} inert doublet, which is $Z_2$-odd, does not develop a Vacuum Expectation Value (VEV) and -- by construction -- does not couple to fermions, plus {1} active $Z_2$-even Higgs doublet, which has a non-zero VEV and couples to fermions in the same way as the SM Higgs doublet. Therefore we shall also refer to the IDM as the I(1+1)HDM to explicitly show the number of inert (I) and active Higgs (H) doublets. An important feature of this model is that, due to the unbroken $Z_2$ symmetry, the lightest neutral $Z_2$-odd particle, coming from the inert doublet, is stable and a suitable DM candidate. 

The I(1+1)HDM, despite being severely constrained by data, remains a viable model for a scalar DM candidate (see the latest analyses, e.g., in \cite{Ilnicka:2015jba,Belyaev:2016lok,Belyaev:2018ext,Kalinowski:2018ylg}). 
This model, by construction, can not contain CP-violation: due to the presence of an exact $Z_2$ symmetry, all parameters in the potential are real. In fact, accommodating CP-violation in multi-inert models requires at least three scalar $SU(2)$ doublets, leading to a 3-Higgs Doublet Model (3HDM). 
Here, one can have two possibilities.
\begin{itemize}
\item 
I(1+2)HDM: a 3HDM with {1} inert doublet plus {2} active Higgs doublets,
\item 
I(2+1)HDM: a 3HDM with {2} inert doublets plus {1} active Higgs doublet.
\end{itemize} 
In the I(1+2)HDM, the inert sector is identical to that of the I(1+1)HDM and CP-violation is introduced in the extended active sector \cite{Grzadkowski:2009bt, Osland:2013sla}. Therefore, the amount of CP-violation is restricted by SM Higgs data, as the Higgs particle observed at the LHC is very SM-like, and by contributions to the Electric Dipole Moments (EDMs) of electron and neutron.

In the I(2+1)HDM, in contrast, the active sector is by construction SM-like, with tree-level interactions identical to those of the SM Higgs, with the exception of possible Higgs decays to new states provided they are sufficiently light\footnote{At loop level, additional states may contribute to Higgs interactions, such as in the $h\to gg, \gamma \gamma $ and $Z\gamma$.}. Here, the inert sector is extended and now contains six new particles, four neutral and two charged ones, i.e., twice as many inert particles as in the I(1+1)HDM. As a result, even without introducing CP-violation, the I(2+1)HDM provides new  coannihilation channels for the DM candidate and revives regions of parameter space that are excluded in the I(1+1)HDM \cite{Keus:2014jha,Keus:2015xya}. 
With the introduction of CP-violation in the inert sector, the neutral inert particles will have a mixed CP quantum number. Note that the \textit{inert} sector is protected by a conserved $Z_2$ symmetry from coupling to the SM particles, therefore, the amount of CP-violation introduced here is not constrained by EDM data. The DM candidate, in this scenario, is the lightest state amongst the CP-mixed inert states which enlivens yet another region of viable DM mass range, with respect to both I(1+1)HDM and CP-conserving I(2+1)HDM \cite{Cordero-Cid:2016krd}.

In this paper, we study electron-positron collider signatures of a CP-violating I(2+1)HDM  via  the process $e^+e^- \to Z^* \to S_iS_j$ ($i,j=1, ... 4$), which has six possible final states, $S_1S_{2,3,4}$, $S_2S_{3,4}$, $S_3S_4$ in the CP-violating case, in comparison to four possible final states, $H_1 A_{1,2}, H_2A_{1,2}$ in the CP-conserving case, wherein $H_{1,2}$ and $A_{1,2}$ have opposite CP-parity\footnote{Recall that as the inert states do not couple to fermions, it is not possible to identify the individual properties of these states.}. Hence, a simple collider energy scan combined with a trivial counting experiment in the detectors revealing six thresholds rather than four will be a clear evidence of CP-violation, whether or not such $S_i$ states will have been previously discovered\footnote{Clearly, also  $S^+_i S^-_j$ ($i,j=1,2$) final states are possible, but these are not discriminatory here, as three thresholds would appear in both cases of CP-conservation and CP-violation. Hence, we will not discuss these here.}. 
This signal by itself does not provide a conclusive evidence for CP-violation, as the observable is a CP-even quantity. However, provided we have observed other processes hinting at a 3HDM, as in our previous and upcoming publications \cite{Keus:2014jha,Keus:2014isa,Keus:2015xya,Cordero-Cid:2016krd,Cordero:2017owj,Preparation}, this signal will eventually
help to 
distinguish between a CP-violating and a CP-conserving 3HDM by considering the
number of observable $Z$ boson decay thresholds in the pair production of neutral scalar
states.
In order to study this phenomenology, we  
 provide several Benchmark Points (BPs), in agreement with all experimental and theoretical bounds, for which we show that the cross section of the $e^+e^- \to Z^* \to S_iS_j$ process could be as large as a few picobarns at $\sqrt{s}$ values accessible by future $e^+e^-$ colliders. The proximity (or otherwise) of these thresholds would serve as characteristic signatures of different BPs with different DM properties.

The layout of the remainder of this paper is as follows. In section \ref{scalar-potential}, we present the details of the scalar potential and the theoretical and experimental limits on its parameters. In section \ref{selection}, we construct and justify our BPs. In section \ref{results}, we show the production cross sections and decay thresholds in our BPs. In section \ref{conclusion}, we conclude and present the outlook for our future studies.

\section{The scalar sector of the I(2+1)HDM}
\label{scalar-potential}

A 3HDM potential symmetric under a group $G$ of phase rotations can be divided into two parts: a phase invariant part, $V_0$, and a collection of extra terms ensuring the symmetry group $G$, $V_G$ \cite{Ivanov:2011ae,Keus:2013hya}. Here, we consider a $Z_2$-symmetry, under which the three Higgs doublets $\phi_{1,2,3}$ transform, respectively, as: 
\be 
\label{generator}
g_{Z_2}=  \mathrm{\rm diag}\left(-1, -1, 1 \right). 
\ee
The resulting potential is of the following form\footnote{Note that adding extra $Z_2$-respecting terms such as
$ 
(\phi_3^\dagger\phi_1)(\phi_2^\dagger\phi_3), 
(\phi_1^\dagger\phi_2)(\phi_3^\dagger\phi_3), 
(\phi_1^\dagger\phi_2)(\phi_1^\dagger\phi_1), 
(\phi_1^\dagger\phi_2)(\phi_2^\dagger\phi_2),
$
does not change the phenomenology of the model. The coefficients of these terms, therefore, have been set to zero for simplicity.}:
\bea
\label{V0-3HDM}
V_{3HDM}&=&V_0+V_{Z_2}, \\
V_0 &=& - \mu^2_{1} (\phi_1^\dagger \phi_1) -\mu^2_2 (\phi_2^\dagger \phi_2) - \mu^2_3(\phi_3^\dagger \phi_3) \nonumber\\
&&+ \lambda_{11} (\phi_1^\dagger \phi_1)^2+ \lambda_{22} (\phi_2^\dagger \phi_2)^2  + \lambda_{33} (\phi_3^\dagger \phi_3)^2 \nonumber\\
&& + \lambda_{12}  (\phi_1^\dagger \phi_1)(\phi_2^\dagger \phi_2)
 + \lambda_{23}  (\phi_2^\dagger \phi_2)(\phi_3^\dagger \phi_3) + \lambda_{31} (\phi_3^\dagger \phi_3)(\phi_1^\dagger \phi_1) \nonumber\\
&& + \lambda'_{12} (\phi_1^\dagger \phi_2)(\phi_2^\dagger \phi_1) 
 + \lambda'_{23} (\phi_2^\dagger \phi_3)(\phi_3^\dagger \phi_2) + \lambda'_{31} (\phi_3^\dagger \phi_1)(\phi_1^\dagger \phi_3),  \nonumber\\
 V_{Z_2} &=& -\mu^2_{12}(\phi_1^\dagger\phi_2)+  \lambda_{1}(\phi_1^\dagger\phi_2)^2 + \lambda_2(\phi_2^\dagger\phi_3)^2 + \lambda_3(\phi_3^\dagger\phi_1)^2  + h.c. \nonumber
\eea
The parameters of $V_0$ are by construction real. We allow for the parameters of $V_{Z_2}$ to be complex, hence introducing explicit CP-violation in the model.

The doublets are defined as
\be 
\phi_1= \doublet{$\begin{scriptsize}$ H^+_1 $\end{scriptsize}$}{\frac{H_1+iA_1}{\sqrt{2}}},\quad 
\phi_2= \doublet{$\begin{scriptsize}$ H^+_2 $\end{scriptsize}$}{\frac{H_2+iA_2}{\sqrt{2}}}, \quad 
\phi_3= \doublet{$\begin{scriptsize}$ G^+ $\end{scriptsize}$}{\frac{v+h+iG^0}{\sqrt{2}}}, 
\label{explicit-fields}
\ee
where $\phi_1$ and $\phi_2$ are the two \textit{inert} doublets, $\langle \phi_1 \rangle = \langle \phi_2 \rangle =0$, while $\phi_3$ is the one \textit{active} doublet, $\langle \phi_3 \rangle =v/$\begin{scriptsize}$ \sqrt{2} $\end{scriptsize} $ \neq 0$, and plays the role of the SM Higgs doublet, with $h$ being the SM Higgs boson and $G^\pm,~ G^0$ the would-be Goldstone bosons.

We assign $Z_2$ charges to each doublet according to the $Z_2$ generator in eq.(\ref{generator}): odd-$Z_2$ charge to the inert doublets, $\phi_1$ and $\phi_2$, and even-$Z_2$ charge to the active doublet, $\phi_3$. It is clear that the symmetry of the potential is respected by the vacuum alignment $(0,0,v/$\begin{scriptsize}$ \sqrt{2} $\end{scriptsize}$)$. 
To make sure that the entire Lagrangian and not only the scalar potential is $Z_2$ symmetric, we assign an even $Z_2$ parity to all SM particles, identical to the active doublet $\phi_3$. With this parity assignment Flavour Changing Neutral Currents (FCNCs) are avoided as the extra doublets are forbidden to couple to fermions and, as dictated by the $Z_2$ symmetry, $\phi_3$ is the only doublet that couples to the fermions though Yukawa interactions identical to those in the SM Yukawa Lagrangian: 
\bea 
\mathcal{L}_{\rm Yukawa} &=& \Gamma^u_{mn} \bar{q}_{m,L} \tilde{\phi}_3 u_{n,R} + \Gamma^d_{mn} \bar{q}_{m,L} \phi_3 d_{n,R} \nonumber\\
&& +  \Gamma^e_{mn} \bar{l}_{m,L} \phi_3 e_{n,R} + \Gamma^{\nu}_{mn} \bar{l}_{m,L} \tilde{\phi}_3 {\nu}_{n,R} + h.c.  
\eea
Here, $\Gamma^{u,d,e,\nu }_{mn}$ are the dimensionless Yukawa couplings for the family indices $m,n$
and $u,d,e,\nu$ labels refer to the SM fermions  in the usual notation.

Note that the scalar $h$ contained in the doublet $\phi_3$ in our model  has the tree-level couplings of the SM Higgs boson. Thus CP-violation is only introduced in the \textit{inert} sector which is forbidden from mixing with the \textit{active} sector by the $Z_2$ symmetry, so that the amount of CP-violation is not limited by EDMs. The lightest amongst the neutral fields from the inert doublets, which now have a mixed CP-charge, $S_1, S_2, S_3, S_4$, is the DM candidate, indeed stable due to the unbroken $Z_2$ symmetry. (We avoid regions of parameter space where one of the charged inert scalars is the lightest.)

\subsubsection*{The parameters of the potential}

The parameters of the potential can be divided into the following categories.
\begin{itemize}
\item \textbf{The Higgs sector parameters}\\
$\mu^2_3, \lambda_{33}$ are Higgs field parameters, fixed by the Higgs mass. We use the value $125$ GeV for the latter and from extremum conditions we have:
\be 
m^2_h = 2\mu^2_3 = 2\lambda_{33} v^2.
\ee

\item \textbf{The dark sector parameters}\\
$\lambda_1, \lambda_{11},\lambda_{22},\lambda_{12}, \lambda'_{12}$ are  inert/dark sector  parameters (inert scalars self-interactions) and in  tree-level analysis they are only constrained through perturbative unitarity and positivity of $V$. Apart from that, they do not play any role in our analysis, as they do not influence tree-level DM and collider phenomenology. We therefore set them to a fixed value of $0.1$. 

\item \textbf{The phenomenologically relevant parameters}\\
$\mu^2_{1},\mu^2_{2},\mu^2_{12}, \lambda_{31},\lambda_{23},\lambda'_{31},\lambda'_{23},\lambda_{2}, \lambda_{3}$ are related to masses of inert scalars and their couplings with the visible sector. These 9 parameters can in principle be determined by independent masses, mixing angles or couplings and the ranges that we allow for them in our numerical studies are
\bea
&& -10~ \mbox{ TeV}^2 < \mu^2_{1},\mu^2_{2},\mu^2_{12} < 10~ \mbox {TeV}^2, \nonumber \\
&& -0.5 < \lambda_{31},\lambda_{23},\lambda'_{31},\lambda'_{23},\lambda_{2}, \lambda_{3} < 0.5 \,. \label{lambdas}
\eea

The only parameters here that can be complex are $\mu^2_{12}$, $\lambda_2$ and $\lambda_3$ for which we use the following notation
\bea 
\label{notation}
&&\mu^2_{12} = \Re \mu^2_{12} +i  \Im\mu^2_{12} = |\mu^2_{12}| e^{i \theta_{12}},\nonumber\\
&&\lambda_2 = \Re \lambda_2 +i  \Im\lambda_2 = |\lambda_2| e^{i \theta_2},
\\
&&\lambda_3 = \Re \lambda_3 +i  \Im\lambda_3 = |\lambda_3| e^{i \theta_3}.
\nonumber
\eea

Note that the phase of $\mu^2_{12}$ is non-physical and can be rotated away with the following redefinition of doublets
\bea 
\phi_1 \to \phi_1 e^{i \theta_{12}/2} ~ \hspace{2mm} ~
& & ~
|\mu^2_{12}| e^{i \theta_{12}} \to |\mu^2_{12}| ,
\nonumber\\
\phi_2 \to \phi_2 e^{-i \theta_{12}/2} ~ ~
& \Longrightarrow & ~
|\lambda_2| e^{i \theta_2} \to |\lambda_2| e^{i (\theta_2+\theta_{12})},
\\
\phi_3 \to \phi_3 ~ \hspace{13mm} ~
& & ~
|\lambda_3| e^{i \theta_3} \to |\lambda_3| e^{i (\theta_3+\theta_{12})}.
\nonumber
\eea
We, therefore, set $\theta_{12}$ to zero for simplicity.

\end{itemize}

\subsubsection*{The \textit{dark democracy} limit}

In our previous papers \cite{Keus:2014jha,Keus:2015xya,Cordero-Cid:2016krd}, we studied a simplified version of the I(2+1)HDM by imposing the following equalities
\be 
\mu^2_1 =\mu^2_2 , \quad \lambda_3=\lambda_2 , \quad \lambda_{31}=\lambda_{23} ,\quad \lambda'_{31}=\lambda'_{23} ,
\ee
which is sometimes referred to as the \textit{dark democracy} limit. After imposing this limit, the model is still explicitly CP-violating when $(\lambda_{22}- \lambda_{11} )
\left[\lambda_1 ({\mu^2_{12}}^*)^2-\lambda^*_{1}(\mu^2_{12})^2 \right] \neq 0$ \cite{Haber:2006ue,Haber:2015pua}. Note that, after rotating away the phase of $\mu_{12}^2$, the amount of CP violation is directly related to the dark sector through the parameters $\lambda_{11,22}$ and a complex $\lambda_1$. 

\subsubsection*{The \textit{dark hierarchy} limit}
In this paper, we study the more general case of the \textit{dark hierarchy}:
\be 
\mu^2_1 =n\mu^2_2 , \quad \textrm{Re}\lambda_3=n \textrm{Re}\lambda_2 , \quad \textrm{Im}\lambda_3=n \textrm{Im}\lambda_2 ,\quad \lambda_{31}=n \lambda_{23} ,\quad \lambda'_{31}=n \lambda'_{23} ,
\ee
where we introduce the dark hierarchy parameter $n$, which can change between $0 \leq n \leq 1$. Boundary values reduce the model to the well-known I(1+1)HDM for $n=0$ and to the dark democracy case for $n=1$. The case of $n>1$ corresponds to a redefinition of states and does not lead to any different phenomenology. 

After imposing the dark hierarchy limit, the only two relevant complex  parameters, $\lambda_2$ and $\lambda_3$, are related through  
$|\lambda_3 | = n |\lambda_2|$ and $ \theta_3 = \theta_2$.
The angle $\theta_2$ is therefore the only relevant CP-violating phase and is referred to as $\theta_{\rm CPV}$ throughout the paper.

Note that $n$ is a discrete parameter which we take as an input in our analysis. In reality, this quantity is subject to RGE effects meaning that our set-up may need readjustment in higher orders. Nonetheless, our low energy phenomenological analysis is always attainable by smooth changes of the $n$ parameter.

\subsection{Physical scalar states}
\label{minimization}

The $Z_2$-conserving minimum of the potential sits at the point $(0,0,\frac{v}{\sqrt{2}})$ with
$ v^2= \frac{\mu^2_3}{\lambda_{33}}$. The resulting mass spectrum of the scalar particles is as follows.

\subsubsection*{The fields from the active doublet}

The fields from the third doublet, $G^0,G^\pm,h$, which play the role of the SM Higgs doublet fields have squared masses of
\bea 
&& m^2_{G^0}= m^2_{G^\pm}=0, \nonumber\\
&& m^2_{h}= 2\mu_3^2 =2\lambda_{33} v^2. 
\eea

\subsubsection*{The charged inert fields}

The two physical charged states, $S_{1}^{\pm}$ and $S_{2}^{\pm}$, from the inert doublets are the eigenstates of the matrix
\be 
\mathcal{M}_C= \left( \begin{array} {cc}
-n \mu_{2 }^{2} + \frac{n}{2} \lambda_{23} v^{2} 
& -\mu_{12}^{2}  \\
- \mu_{12}^{2}   &  -\mu_{2}^{2} + \frac{1}{2} \lambda_{23} v^{2}  \end{array} \right) ,
\ee
with eigenvalues:
\be 
m^2_{S^\pm_{1,2}}
=
\frac{1}{4} \left((n+1)(-2 \mu_2^2+\lambda_{23}v^2)\; \mp \;\sqrt{16 (\mu_{12}^2)^2+(n-1)^2 \left(\lambda_{23} v^2-2 \mu_2^2\right)^2}\right).
\ee
In terms of gauge states from eq.(\ref{explicit-fields}) $S^\pm_i$ are defined through:
\be 
\left( \begin{array} {c}
S_1^\pm\\
S_2^\pm \end{array} \right )= 
\left( \begin{array} {cc}
\cos \alpha_c & \sin \alpha_c \\
-\sin \alpha_c & \cos \alpha_c 
\end{array} \right )
\left( \begin{array} {cccc}
H_1^\pm\\
H_2^\pm \end{array} \right )
\quad 
\mbox{with} 
\quad 
\tan2\alpha_c = \frac{2 \mu_{12}^2}{(n-1) (\mu_2^2 - \lambda_{23} v^2/2)}.
\ee 
We require $\pi/2 < \alpha_c < \pi$, so that $m_{S_1^\pm} < m_{S_2^\pm}$. 

\subsubsection*{The  neutral inert fields}

The neutral mass-squared matrix  in the $(H_1,H_2,A_1,A_2)$ basis is
\be 
\mathcal{M}_N= \frac{1}{4}\left( \begin{array} {cccc}
n \;\Lambda^+_c
& -2\mu^2_{12}
& -n \;\Lambda_s
& 0 
\\[2mm]
-2\mu^2_{12}
& \Lambda^+_c
& 0 
& \Lambda_s 
\\[2mm]
-n \;\Lambda_s
& 0 
&  n \;\Lambda^-_c
& -2\mu^2_{12} 
\\[2mm]
0 
& \Lambda_s
& -2\mu^2_{12}
&  \Lambda^-_c 
\end{array} \right ),
\label{neutral-mass-squared}
\ee
with
\be  
\Lambda_s=  2\lambda_2 \sin \theta_{\rm CPV} v^2
\quad
\mbox{and}
\quad 
\Lambda^\pm_c =-2\mu^2_2 +
(\lambda_{23}+\lambda'_{23} \pm 2 \lambda _2  \cos \theta_{\rm CPV}) v^2.
\ee  
Note that, in the CP-conserving limit, $\theta_{\rm CPV} = 0,\pi$ leads to $\Lambda_s=0$ which reduces $\mathcal{M}_N$ to a block diagonal matrix with no mixing between the states with opposite CP-parity, namely between $H_{1,2}$ and $A_{1,2}$. 

We diagonalise the neutral mass-squared matrix numerically, $\mathcal{M}_N^{\rm diag} = R^T \mathcal{M}_N R$, to derive our mass eigenstates, $S_i$, in terms of the gauge eigenstates in eq.(\ref{explicit-fields}), 
\be
\left( \begin{array} {cccc}
S_1\\
S_2 \\
S_3 \\
S_4  \end{array} \right )= R_{ij}
\left( \begin{array} {cccc}
H_1\\
H_2\\
A_1 \\
A_2  \end{array} \right ).
\ee
We adopt a notation where $m_{S_1} < m_{S_2}  <m_{S_3} <m_{S_4}$, hence choosing $S_1$ as DM candidate. We use 
\be 
\label{parameters}
|\mu^2_{12}| , \;\lambda_{23}, \; \lambda'_{23}, \; \mu^2_2, \; \lambda_2, \; \theta_{\rm CPV}, n
\ee 
as the set of input parameters to define our BPs in a forthcoming section.

\subsection{Constraints on the parameters}\label{constraints}

In this section, we discuss the latest theoretical and experimental constrains that are applicable to our studies. The I(2+1)HDM is a model which is partially already within reach of current collider as well as  DM experiments, and their results constrain parts of parameter space.

\subsubsection*{Theoretical constraints}
In the ``dark hierarchy'' limit, theoretical requirements of boundedness of the potential and positive-definiteness of the Hessian are taken into account. All couplings fulfil perturbative unitarity limits, i.e., they  take absolute values $\lambda_i\leq\,4\,\pi$, as noted in eq.(\ref{lambdas}). For detailed formulas see \cite{Keus:2014jha,Keus:2015xya,Cordero-Cid:2016krd}.

\subsubsection*{Experimental constraints} 
\begin{itemize}
\item Higgs decays and signal strengths\\
The latest measurement of the SM-like Higgs boson's width gives $\Gamma_\text{tot}\,=3.2^{+2.8}_{-2.2}$ MeV, with 95\% CL upper limit of  $\Gamma_\text{tot} \leq 9.16$ MeV \cite{Sirunyan:2019twz}. In our model, the total width of the SM-like Higgs boson can be modified through two mechanisms. If inert scalars are light, $m_{S_i} <m_h/2$, we can expect a measurable contribution to Higgs invisible decays. This sets strong limits on the Higgs-inert couplings in the light mass region. Furthermore, the partial decay $\Gamma(h\to \gamma\gamma)$ can be significantly changed through the two charged inert scalar contributions, as new physics corrections are formally of the same order as the SM process. In this work we use the combined  ATLAS and CMS Run I limit for the signal strength $h\to \gamma \gamma$, $\mu_{\gamma \gamma} = 1.14^{+0.38}_{-0.36}$ \cite{Khachatryan:2016vau}\footnote{In Run II, ATLAS reports $\mu_{\gamma \gamma} = 0.99^{+0.14}_{-0.14}$ \cite{Aaboud:2018xdt}, and CMS reports  $\mu_{\gamma \gamma} = 1.18^{+0.17}_{-0.14}$ \cite{Sirunyan:2018ouh}. Our BPs are within $1\sigma$ agreement with ATLAS and $2\sigma$ agreement with CMS results.}. By construction, the Higgs particle is SM-like and couplings to gluons, massive gauge bosons and fermions are equal to the SM values.

The aim of this paper is to present a way of testing the model at future linear colliders. However, before either CLIC or ILC are built, the analysis of current and future runs at the LHC are going to provide stronger constraints on the parameter space. Therefore, it is also important to establish if the model is in agreement with projected LHC exclusion limits. The diphoton signal strengths $\mu$ for benchmarks A, B, C (presented in details in following sections) are $\mu_A = \mu_B = 0.937, \mu_C = 0.853$, respectively. They are all within the 3$\sigma$ limit of current measurement. Note, that the signal strength is the same for two benchmark A and B, which correspond to a vastly different dark sector. This further stresses the importance of using linear colliders to fully explore the models with extended scalar sector. Assuming predictions from the HL-LHC \cite{1902.00134}, we expect $\mu=1.00 \pm 0.04$ in the ggF channel, and $\mu=1.00^{+ 0.10}_{-0.09}$ in the VBF channel, which means these benchmarks will still be in agreement with the observation.

To be in agreement with current experimental constraints we have to minimize the contribution of new particles to Higgs decays. Two benchmarks have particles with masses below $m_h/2$, benchmarks B and C. However, as $\sum_{i,j} BR(h\to S_i S_j)_B = 0.07\%, \sum_{i,j} BR(h\to S_i S_j)_C = 0.5\%, i,j = 1,2,3,4$, we do not observe any significant changes in the total decay width of the Higgs with respect to the SM prediction. Furthermore, even taking into account the most promising estimation of the final constraint on Higgs invisible decays from HL-LHC and LeHC, $\lesssim 4-7\%$ and $\lesssim 2.25\%$, our chosen benchmarks will still be in agreement with the observation.

In short, projected sensitivity limits from Higgs physics of future runs of the LHC are not going to provide any relevant constraints for the parameter range of the model considered in this paper.

\item Gauge bosons widths\\
Similarly to the Higgs width, if new particles are sufficiently light, they could significantly change the total width of Electro-Weak (EW) gauge bosons. We control this by forbidding decays $W^\pm \to S_i S_j^\pm$ and $Z\to S_i S_j,S_i^+S_j^-$ through enforcing:
\be 
\label{eq:gwgz}
m_{S_i}+m_{S^\pm_i}\,\geq\,m_W^\pm,~~
\,m_{S_i}+m_{S_j}\,\geq\,m_Z,\,~~
2\,m_{S_i^\pm}\,\geq\,m_Z.
\ee

\item EW Precision Observables (EWPOs)\\ 
We require  a $2\sigma$, i.e., a $95 \%$ Confidence Level (CL), agreement, parameterised through the EW oblique parameters $S,T,U$ \cite{Altarelli:1990zd,Peskin:1990zt,Peskin:1991sw,Maksymyk:1993zm}.
{Just like in the 2HDM, it suffices here to have in the dark sector (near) degeneracy between each charged state and one or two of the neutral ones, condition which  is satisfied by all our BPs.}

\item Charged scalar mass and lifetime\\ 
We take a conservative lower estimate on the masses of charged scalars \cite{Pierce:2007ut} $m_{S^\pm_i} > 70$ GeV ($i=1,2$). We also ensure that neither of these particles are quasi-stable by setting a limit for the charged scalar lifetime to be $\tau\,\leq\,10^{-7}\,{\rm s}$  \cite{Heisig:2018kfq}.

\item Searches for new particles at colliders\\
As in previous works of ours, we use LEP 2 searches for supersymmetric particles (chiefly, sneutrinos and sleptons) re-interpreted for the IDM in order to exclude the region of masses where the following condition are simultaneously satisfied \cite{Lundstrom:2008ai} ($i,j=2, ... 4$):
\be 
\label{eq:leprec}
m_{S_i}\,\leq\,100\,\gev,\,~~
m_{S_1}\,\leq\,80\,\gev,\,\, ~~
\Delta m {(S_i,S_1)}\,\geq\,8\,\gev,
\ee
since this would lead to a visible di-jet or di-lepton signal\footnote{Effects of CP-violation directly onto the $ZS_iS_j$ vertex are marginal in our BPs.}.

Benchmarks are in agreement with null-results for additional neutral scalar searches at the LHC, where we make use of HiggsBounds-5.4.0beta \cite{0811.4169,1102.1898,1311.0055,1507.06706} and HiggsSignals-2.2.3beta \cite{1305.1933}. 
This follows the analysis performed in \cite{Kalinowski:2018ylg} for the I(1+1)HDM, which is a model with similar signatures to one studied in this paper, especially for benchmark A. 

One of possible ways of testing the model would be using searches for multilepton final states with missing transverse energy. However, current strategy at the LHC uses a relatively large cut on missing transverse energy, which corresponds to a rather large mass splittings between scalars in the dark sector, and that reduces the production cross sections. Benchmarks with smaller mass splittings between scalars have large enough cross-section to be produced in abundance even at the current stage of the LHC, however they require smaller cuts on missing energy to be detected. With current analysis setup used at the LHC, the considered parameter range of the I(2+1)HDM model will remain invisible even for future HL-LHC or HE-LHC, especially for benchmarks of type B and C. We strongly encourage the LHC experimental collaborations to expand their search region in multilepton final states and missing $E_T$ by allowing for lower cuts on missing energy, which would give us access to the considered parameter space of the model. Note that mono-photon searches at LEP may put more limiting constraints on light DM particles with light mediators \cite{Fox:2011fx,Essig:2013vha}. However, in our chosen benchmark points with $m_{DM} \geq 50$ GeV and mediator mass of $m_Z = 91$ GeV, direct detection is the more constraining probe of the model.

\item DM measurements\\
We require  agreement with  relic density limits from the Planck experiment \cite{Aghanim:2018eyx}:
\begin{equation}\label{eq:planck}
\Omega_c\,h^2\,=\,0.120\,\pm\,0.001,
\end{equation}
as well as with  the latest XENON1T results for direct DM searches \cite{Aprile:2018dbl}. In the region of masses we are considering in this paper, indirect detection experiments (e.g., FermiLAT)  do not place any additional constraints upon the parameter space. 

Note that, due to the absence of any $S_iS_i Z$ vertex, the only loop-induced process that could potentially contribute to the DM-nucleon interaction is the rightmost diagram in Fig.~\ref{DD-fig}. However, since direct detection experiments are performed at the zero-momentum limit, for this diagram to contribute to the DM-nucleon cross-section, the mass splitting between $S_1$ and the next lightest inert scalar, $S_{2,3,4}$, has to be of the keV order. In all our benchmark points, the $S_i-S_j$ mass splitting is of above 1 GeV. Therefore, the only tree-level contribution to the DM-nucleon cross-section is from the $S_1N\to S_1N$ process mediated by the Higgs boson.

The one-loop contributions, represented\footnote{See figure 1 in \cite{Klasen:2013btp} for a full list of such diagrams.} by the rightmost diagram in Fig.~\ref{DD-fig}, are shown to have a relevant contribution in the I(1+1)HDM \cite{Klasen:2013btp}, which is CP conserving by construction. However, due to CP violation in our model, the strength of the gauge couplings is reduced. Therefore, the results of \cite{Klasen:2013btp} are not directly applicable here. We postpone the detailed study of DM direct detection one-loop contributions to a future publication, since the focus of this paper is on the collider signatures of the model.
\end{itemize}

\begin{figure}
\begin{center}
\begin{minipage}{0.3\linewidth}  
\centering       
\begin{tikzpicture}[thick,scale=1.0]
\draw[dashed] (8.5,-0.2) -- node[black,above,xshift=-0.8cm,yshift=0.0cm] {$S_1$} (9.5,-0.7);
\draw[dashed] (9.5,-0.7) -- node[black,above,xshift=1.1cm,yshift=0.0cm] {$S_{1,2,3,4}$} (10.5,-0.2);
\draw[dashed] (9.5,-2) -- node[black,above,xshift=-0.3cm,yshift=-0.3cm] {$h$} (9.5,-0.7);
\draw[particle] (8.5,-2.5) -- node[black,above,xshift=-0.8cm,yshift=-0.6cm] {$N$} (9.5,-2);
\draw[particle] (9.5,-2) -- node[black,above,xshift=0.8cm,yshift=-0.6cm] {$N$} (10.5,-2.5);
\end{tikzpicture}
\end{minipage}
\hspace{-5mm}
\begin{minipage}{0.3\linewidth}  
\centering       
\begin{tikzpicture}[thick,scale=1.0]
\draw[dashed] (8.5,-0.2) -- node[black,above,xshift=-0.8cm,yshift=0.0cm] {$S_1$} (9.5,-0.7);
\draw[dashed] (9.5,-0.7) -- node[black,above,xshift=1.1cm,yshift=0.0cm] {$S_{2,3,4}$} (10.5,-0.2);
\draw[photon] (9.5,-2) -- node[black,above,xshift=-0.3cm,yshift=-0.3cm] {$Z$} (9.5,-0.7);
\draw[particle] (8.5,-2.5) -- node[black,above,xshift=-0.8cm,yshift=-0.6cm] {$N$} (9.5,-2);
\draw[particle] (9.5,-2) -- node[black,above,xshift=0.8cm,yshift=-0.6cm] {$N$} (10.5,-2.5);
\end{tikzpicture}
\end{minipage}
\hspace{-5mm}
\begin{minipage}{0.3\linewidth}  
\centering       
\begin{tikzpicture}[thick,scale=1.0]
\draw[dashed] (7.9,0.5) -- node[black,above,xshift=-0.8cm,yshift=0.0cm] {$S_1$} (8.9,0);
\draw[dashed] (8.9,0) -- node[black,above,xshift=0.0cm,yshift=0.0cm] {$S_{2,3,4}$} (10.1,0);
\draw[dashed] (10.1,0) -- node[black,above,xshift=0.8cm,yshift=0.0cm] {$S_1$} (11.1,0.5);
\draw[photon] (9.5,-1) -- node[black,above,xshift=-0.3cm,yshift=-0.4cm] {$Z$} (8.9,0);
\draw[photon] (9.5,-1) -- node[black,above,xshift=0.3cm,yshift=-0.4cm] {$Z$} (10.1,0);
\draw[dashed] (9.5,-1) -- node[black,above,xshift=-0.3cm,yshift=-0.3cm] {$h$} (9.5,-2);
\draw[particle] (8.5,-2.5) -- node[black,above,xshift=-0.8cm,yshift=-0.6cm] {$N$} (9.5,-2);
\draw[particle] (9.5,-2) -- node[black,above,xshift=0.8cm,yshift=-0.6cm] {$N$} (10.5,-2.5);
\end{tikzpicture}
\end{minipage}
\caption{The only tree-level contribution to the direct detection cross-section is from the tree-level $S_1N\to S_1N$ process mediated by the Higgs boson (left). With $m_{S_{j}}-m_{S_1} \gg$ few keV ($j=2,3,4$), the $S_1N\to S_jN$ processes do not contribute to the direct detection cross-section. The one-loop processes are represented by the rightmost diagram.}
\label{DD-fig}
\end{center}
\end{figure}
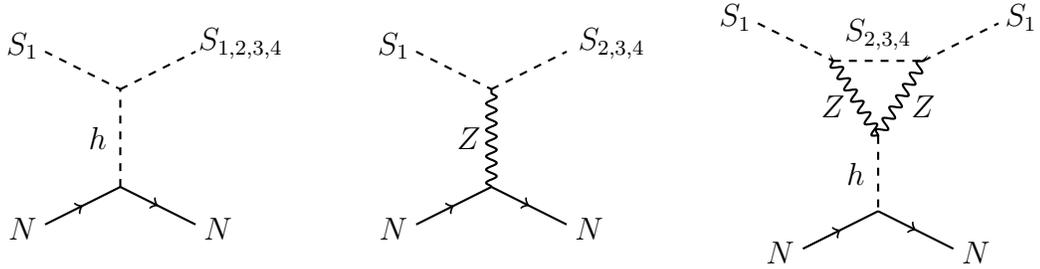

\section{Selection of BPs}\label{selection}

In our previous papers \cite{Keus:2014jha,Keus:2015xya,Cordero-Cid:2016krd}, we discussed DM phenomenology of the I(2+1)HDM in detail, where in addition to standard Higgs/gauge mediated annihilation channels of DM, there exist the possibility of coannihilation with heavier states, provided they are close in mass. This is a feature of models with extended dark sectors and contributes to changes in DM relic density.
It is important to note that the relevance of this effect will depend not only on the DM mass and the mass splittings  but also on the strength of the standard DM annihilation channels. For example, in some BPs presented in later sections, coannihilation, although possible, is responsible for less than 1\% of the overall contributions to $\Omega_{\rm DM}h^2$ because of a very strong DM annihilation into gauge bosons. 

As detailed in \cite{Keus:2014jha,Keus:2015xya,Cordero-Cid:2016krd}, the generic expected behaviour of the I(2+1)HDM in different DM mass regions, is as follows.
\begin{itemize}
\item 
Regions with $m_{\rm DM} \leq 45$ GeV are ruled out due to the $Z$ gauge boson width constraints.
\item 
In the $45$ GeV $ \leq m_{\rm DM} \leq 53$ GeV range, $S_1$ mainly (co)annihilates with SM fermions,
\be 
S_i S_j \to h / Z \to f \bar f.
\label{45-53}
\ee
In this region, the $h\to invisible$ channel is open and requires a very small Higgs-DM coupling to satisfy the experimental bounds.
\item 
In the $53$ GeV $\leq m_{\rm DM} \leq 75$ GeV range, (co)annihilations could also be mediated by the SM gauge bosons, $V=Z, W^\pm$,
\be 
S_i S_j \to V V^* \to V f \bar f, \qquad S_i S_j \to V^* V^* \to f\bar f f \bar f.
\label{53-75}
\ee
The $h\to invisible$ channel is closed, however, strong bounds from direct and indirect detection experiments require a very small Higgs-DM coupling.

\item 
In the $75$ GeV $ \leq m_{\rm DM} \lesssim 375$ GeV range, $S_1$ (co)annihilation with gauge bosons,
\be 
S_i S_j \to h \to V V, \qquad S_i S_j \to V V,
\label{75-375}
\ee
is so strong that the model may fail to provide 100\% of the observed DM relic density (so that a second DM component may need to be invoked, albeit in
a wider framework than our I(2+1)HDM).

\item 
In the $m_{\rm DM} \gtrsim 375$ GeV range, coannihilations with $S_j^\pm$,
\be 
S_i S_j^\pm \to W^\pm \to f f',
\label{375}
\ee
interfere destructively with (co)annihilation to gauge bosons. As a result, the model provides 100\% of the observed DM relic density. 

\end{itemize}

Taking all theoretical and experimental bounds (listed in section \ref{constraints}) into account, we have devised a few benchmark scenarios which show interesting phenomenology. 
In this paper, we do not consider the heavy mass region for the DM candidate ($m_{\rm DM} \gtrsim m_Z$), due to the fact that the $e^+e^-$ production cross section of the heavy inert scalars drops significantly with an increasing $m_{\rm DM}$ value (since $m_{\rm DM}\equiv m_{S_1}<m_{S_2}<m_{S_3}<m_{S_4}$). These points could be tested if $\sqrt{s}$ is increased beyond the maximum value that we will consider, of  500 GeV.  
Also, as the heavy mass region corresponds to a semi-degenerate spectrum (in order to satisfy DM relic density bounds), we are not expecting to see there the interesting signatures and separation of thresholds that can be detected for the medium mass region (45 GeV $\leq m_{\rm DM} \leq m_Z$), as it will be discussed in section \ref{results}.

In such a medium mass region, the I(2+1)HDM provides three distinctive types of benchmark scenarios.
To avoid all exclusion limits, we require a very small Higgs-DM coupling, $g_{h{\rm DM}} \simeq 10^{-3}$.

\begin{itemize}
\item \textbf{Scenario A}\\[2mm]
This is a case with large mass splittings, of order 50 GeV or so, between the DM candidate and all other inert particles:
\be 
m_{S_1} \ll m_{S_2}, m_{S_3}, m_{S_4}, m_{S^\pm_1}, m_{S^\pm_2}.
\ee
In this scenario no coannihilation channels are present and therefore $S_1$ only annihilates through the Higgs boson to other SM particles.

In the $45$ GeV $ \leq m_{\rm DM} < 53$ GeV range, the tiny $g_{h{\rm DM}}$ does not provide an efficient annihilation of DM and is therefore forbidden by relic density observations.
Within the mass range $53$ GeV $ \leq m_{\rm DM} \leq 75 $ GeV,  this scenario could easily accommodate points with a very small $g_{h{\rm DM}}$ and avoid all exclusion limits.  

\item \textbf{Scenario B}\\[2mm]
This is a case 
with a small mass splitting, of order 20\% of $m_{\rm DM}$, between the DM and one inert neutral particle:
\be 
m_{S_1} \sim m_{S_2} \ll m_{S_3},  m_{S_4}, m_{S^\pm_1}, m_{S^\pm_2}.
\ee
In this scenario the DM can coannihilate with its only particle close in mass, $S_2$. This choice also leads to a relatively small mass splitting between $S_3$ and $S_4$, and effectively separating the neutral sector into two groups, with each generation accompanied by a charged scalar. 

In the $45$ GeV $\leq m_{\rm DM} < 53$ GeV range, due to the existence of the coannihilation channel, DM is under-produced. 
In the mass range $53$ GeV $\leq m_{\rm DM} \leq 75$ GeV, where the destructive interference with coannihilation to gauge bosons comes into play, this scenario could accommodate points with very small $g_{h{\rm DM}}$ and 100\% DM contribution.

If one relaxes the re-interpreted Supersymmetric limits, discussed in section \ref{constraints}, and allows for larger mass splittings between $S_1$ and $S_2$, the strength of the $S_1 S_2$ coannihilation channel could be reduced. As a result, this scenario can provide points where $S_1$ contributes to 100\% of DM relic density in the whole $45$ GeV $ \leq m_{\rm DM} \leq 75$ GeV range with very small $g_{h{\rm DM}}$.

%the $hS_1S_2$ coupling is zero, due to the exact CP-even nature of the SM Higgs boson, $h$.

\item \textbf{Scenario C}\\[2mm]
This is a case with all neutral particles close in mass:
\be 
m_{S_1} \sim m_{S_2} \sim m_{S_3} \sim  m_{S_4} \ll m_{S^\pm_1} \sim m_{S^\pm_2}.
\ee
In this scenario the DM can coannihilate with all other neutral inert particles. 
Charged scalars are considerably heavier and do not participate in the coannihilation. 

Across the whole low and medium mass range, this scenario under-produces DM, due to the small mass splittings of the neutral inert particles which in turn strengthens the coannihilation channels.
Contrary to the previous case, this situation cannot be resolved by relaxing the re-interpreted Supersymmetric limits and allowing for larger mass splittings. This is due to the large number of the coannihilation channels. As a result, with a very small $g_{h{\rm DM}}$, this scenario will will not be able to contribute to 100\% of the observed relic density.

\end{itemize}

\section{The $ e^{+} e^{-}  \to  Z^{*} \to S_{i} S_{j}$ cross section}
\label{results}

We calculate the the $ e^{+} e^{-}  \to  Z^{*} \to S_{i} S_{j}$ cross section  at tree-level as \cite{1611.08518,Heinemeyer:2015qbu}
\be
\sigma_{S_{i}S_{j}}= 
\frac{ \pi \; \alpha^{2} \; s \; g_{ZS_iS_j}^2}{24 \; (s-m_{Z}^2)^{2} \; g^2}  ( \frac{ 8 \; {\sin\theta_W}^4 - 4 {\sin\theta_W}^{2} + 1}{{\sin\theta_W}^{4} {\cos\theta_W}^{6}}) f^{3} (x,y),
\ee
with $x= m_{S_i}^{2}/s$, $y=m_{S_j}^{2}/s$ and the function
\begin{eqnarray}
f(x,y)= \sqrt{ 1 + x^{2} + y^{2} - 2x -2y -2xy}.
\end{eqnarray}
The $ZS_iS_j$  couplings are defined according to eq.(\ref{gZSS}). Needless to say,  cross sections with lighter final states will  peak earlier at smaller $\sqrt{s}$) while those with larger $g_{ZSiSj}$ coupling will be  larger.

Following the discussion in section \ref{selection}, we have chosen three representative BPs from each possible scenario in the medium DM mass region. 
For all BPs, we aim to have at least one set of masses with $m_{S_i} + m_{S_j} < 250$ GeV, which should lead to at least one channel being fully accessible at the first stage of a future $e^+e^-$ collider, the so-called `Higgs factory' run.

Below, for each BP, we list the input parameters, i.e., masses of particles and all relevant couplings, following the convention:
\begin{eqnarray} 
&& \mathcal{L}_{\rm gauge} \supset g_{ZS_iS_j} Z_\mu (S_i \partial^\mu S_j - S_j \partial^\mu S_i), \label{gZSS}\\
&& \mathcal{L}_{\rm scalar} \supset \frac{v}{2}g_{S_i S_i h} h S_i^2+ v g_{S_i S_j h} h S_i S_j + v g_{S_i^\pm S_j^\mp h} h S_i^\pm S_j^\mp. \label{ghSS}
\end{eqnarray}

\subsubsection*{Benchmark A}
The input parameters for Benchmark A are defined as
\be 
\begin{array}{c}
n= 0.6, \quad 
\lambda'_{23}= -0.16, \quad  
\lambda_{23}= 0.29, \quad  
\lambda_2= 0.067,\\[1mm]
\theta_{\rm CPV}=15 \pi/16, \quad 
\mu_2^2= -13800~{\rm GeV}^2, \quad  
\mu_{12}^2= 5050~{\rm GeV}^2,
\end{array}
\ee
which lead to the following masses for the dark particles:
\be 
\begin{array}{c}
m_{S_1}=72.331~{\rm GeV}, \quad
m_{S_2}=103.313~{\rm GeV}, \quad
m_{S_1^\pm}=106.235~{\rm GeV},\\[1mm]
m_{S_3}=129.467~{\rm GeV}, \quad
m_{S_4}=155.178~{\rm GeV}, \quad
m_{S_2^\pm}=157.588~{\rm GeV},
\end{array}
\ee
and the following gauge couplings:
\be 
\begin{array}{c}
g_{ZS_1S_2}=0.366, \quad
g_{ZS_1S_3}=0.0397, \quad
g_{ZS_1S_4}=0.0401,\\[1mm]
g_{ZS_2S_3}=0.04006, \quad
g_{ZS_2S_4}=0.0397, \quad
g_{ZS_3S_4}=0.366.
\end{array}
\ee

\begin{figure}[h]
\begin{center}
\includegraphics[scale=0.7]{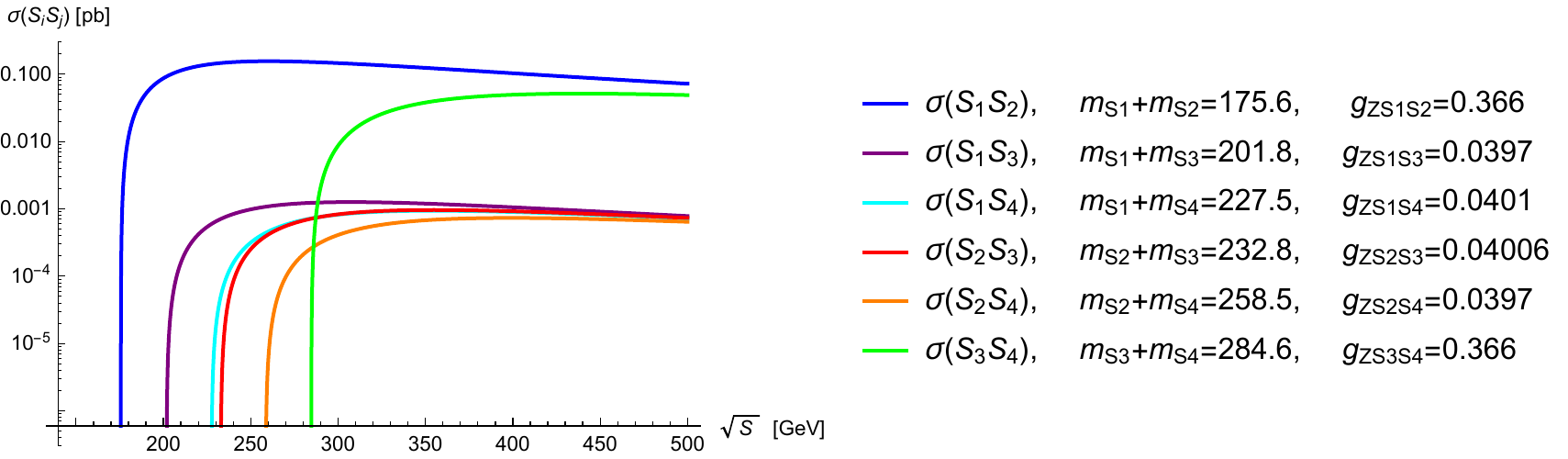}
\caption{The $ e^{+} e^{-}  \to  Z^{*} \to S_{i} S_{j}$ cross section for BP A with masses in GeV. }
\label{BP-A-fig}
\end{center}
\end{figure}

A characteristic signature of type A BPs, as shown in figure \ref{BP-A-fig}, is a pattern of very distinct thresholds that open up as $\sqrt{s}$ increases, all easily resolvable thanks to the fine beam resolution available at future electron-positron machines. 
Here, the lightest and heaviest final states dominate over those with intermediate rest masses since the size of the cross section is dictated by the  $ZS_iS_j$ couplings.

\subsubsection*{Benchmark B}
The input parameters for Benchmark B are defined as
\be 
\begin{array}{c}
n=0.5, \quad 
\lambda'_{23}=-0.145, \quad  
\lambda_{23}=0.171, \quad  
\lambda_2= 0.013,\\[1mm]
\theta_{\rm CPV}=7\pi /8, \quad 
\mu_2^2= -15900~{\rm GeV}^2, \quad  
\mu_{12}^2=7950~{\rm GeV}^2,
\end{array}
\ee
which lead to the following masses for the dark particles:
\be 
\begin{array}{c}
m_{S_1}=55.441~{\rm GeV}, \quad
m_{S_2}=63.219~{\rm GeV},\quad 
m_{S_1^\pm}=79.184~{\rm GeV},\\[1mm]
m_{S_3}=144.377~{\rm GeV},\quad 
m_{S_4}=148.842~{\rm GeV},\quad 
m_{S_2^\pm}=159.203~{\rm GeV},
\end{array}
\ee
and the following gauge couplings:
\be 
\begin{array}{c}
g_{ZS_1S_2}=0.37, \quad
g_{ZS_1S_3}=0.007, \quad
g_{ZS_1S_4}=0.007,\\[1mm]
g_{ZS_2S_3}=0.007, \quad
g_{ZS_2S_4}=0.007, \quad
g_{ZS_3S_4}=0.37.
\end{array}
\ee

\begin{figure}[h]
\begin{center}
\includegraphics[scale=0.7]{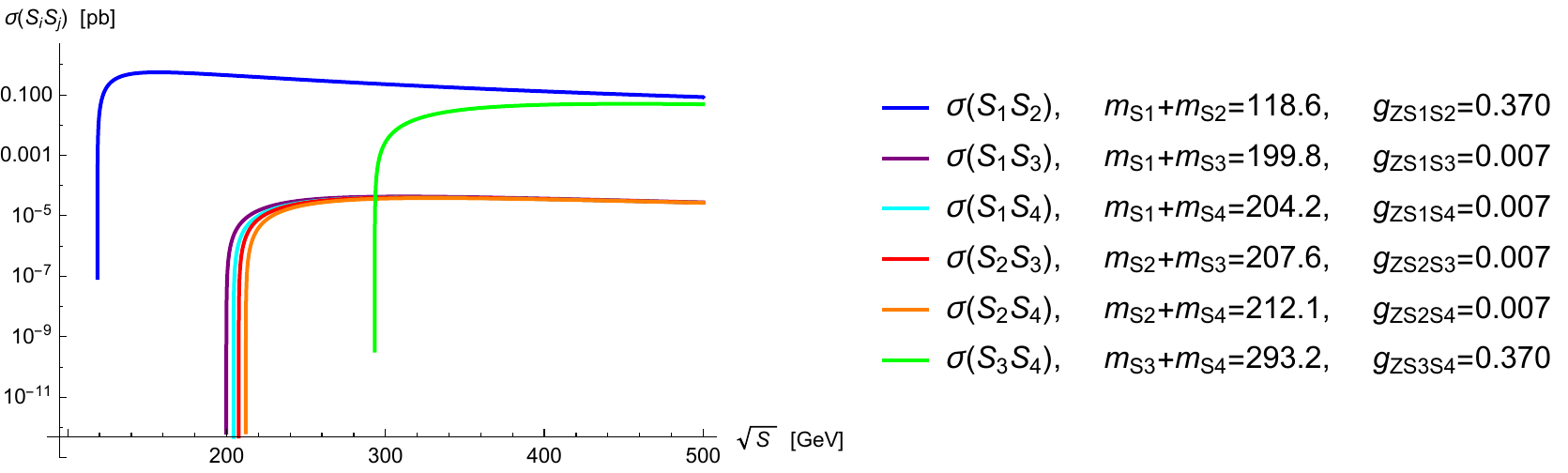}
\caption{The $ e^{+} e^{-}  \to  Z^{*} \to S_{i} S_{j}$ cross section for BP B with masses in GeV. }
\label{BP-B-fig}
\end{center}
\end{figure}

A characteristic signature of type B BPs, as shown in figure \ref{BP-B-fig}, is two distinct thresholds, one (single) at low $\sqrt{s}$ and another (single) at high $\sqrt{s}$, plus several (very closely spaced) ones at mid $\sqrt{s}$. Here too it is the lightest and heaviest final states that dominate over those with intermediate rest masses as dictated by the  $ZS_iS_j$ coupling strengths.

\subsubsection*{Benchmark C}
The input parameters for Benchmark C are defined as
\be 
\begin{array}{c}
n=0.8, \quad 
\lambda'_{23}=-0.295, \quad  
\lambda_{23}=0.294 \quad  
\lambda_2= 0.0009\\[1mm]
\theta_{\rm CPV}=31 \pi /32, \quad 
\mu_2^2= -3400~{\rm GeV}^2, \quad  
\mu_{12}^2=250~{\rm GeV}^2,
\end{array}
\ee
which lead to the following masses for the dark particles:
\be 
\begin{array}{c}
m_{S_1}=50.925~{\rm GeV},\quad 
m_{S_2}=51.793~{\rm GeV},\quad 
m_{S_1^\pm}=99.176~{\rm GeV},\\[1mm]
m_{S_3}=58.555~{\rm GeV}, \quad
m_{S_4}=59.459~{\rm GeV},\quad 
m_{S_2^\pm}=111.136~{\rm GeV},
\end{array}
\ee
and the following gauge couplings:
\be 
\begin{array}{c}
g_{ZS_1S_2}=0.37, \quad
g_{ZS_1S_3}=0.0025, \quad
g_{ZS_1S_4}=0.0028,\\[1mm]
g_{ZS_2S_3}=0.0028, \quad
g_{ZS_2S_4}=0.0025, \quad
g_{ZS_3S_4}=0.37.
\end{array}
\ee

\begin{figure}[h]
\begin{center}
\includegraphics[scale=0.7]{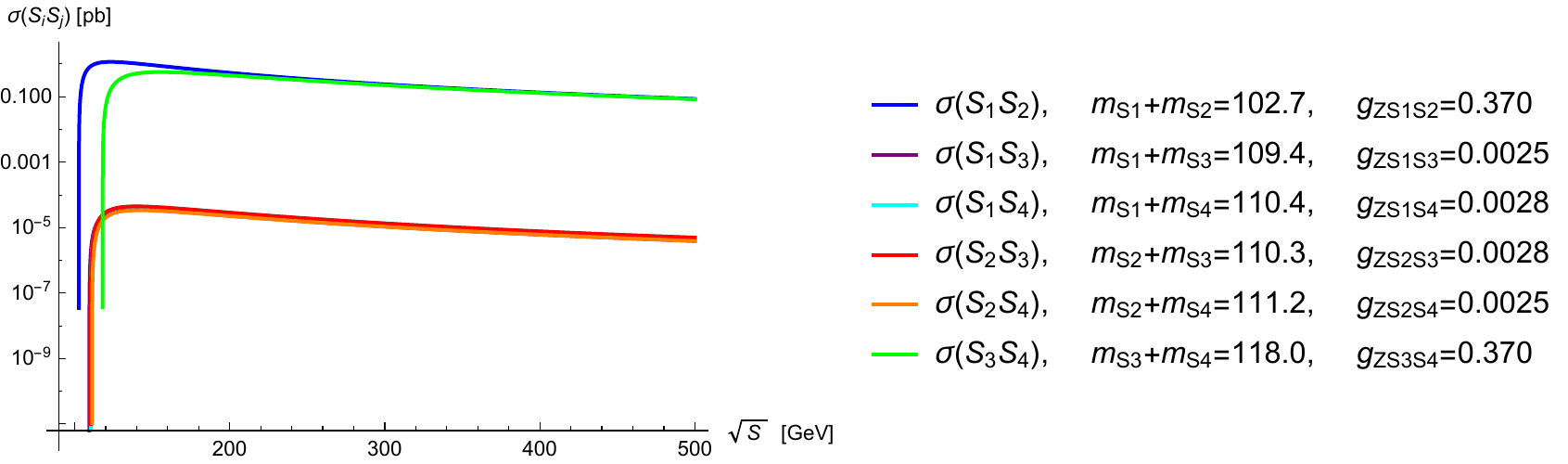}
\caption{The $ e^{+} e^{-}  \to  Z^{*} \to S_{i} S_{j}$ cross section for BP C with masses in GeV. }
\label{BP-C-fig}
\end{center}
\end{figure}

Points from the C type benchmark scenario, as shown in figure \ref{BP-C-fig}, have the specific characteristic of seeing all thresholds (nearly) overlapping at low $\sqrt{s}$ values. The size of the various cross sections is again dictated by the  $ZS_iS_j$ couplings and thus is  largest for $S_1S_2$ and $S_3S_4$ over any of $S_1S_3$, $S_1S_4$, $S_2S_3$ and $S_2S_4$, as previously seen already.

\subsection{Significance of the signal over the SM background}
\label{background}

Linear colliders are perfect machines to test the inert sector of multi-scalar models, e.g. in leptonic or semi-leptonic channels with missing transverse energy. Full analysis is beyond the scope of this work. However, as shown in \cite{Kalinowski:2018kdn} in the context of the IDM, with the use of multi-variate analysis and proper cut selection it is possible to claim 5$\sigma$ discovery for $\sqrt{s} = 500$ GeV and integrated luminosity of 1 ab$^{-1}$ if the sum of masses of the neutral scalar pair is below 330 GeV \cite{Zarnecki:2019poj}, which all of our proposed benchmarks fulfil.

\begin{figure}[h]
\begin{center}
\includegraphics[scale=0.75]{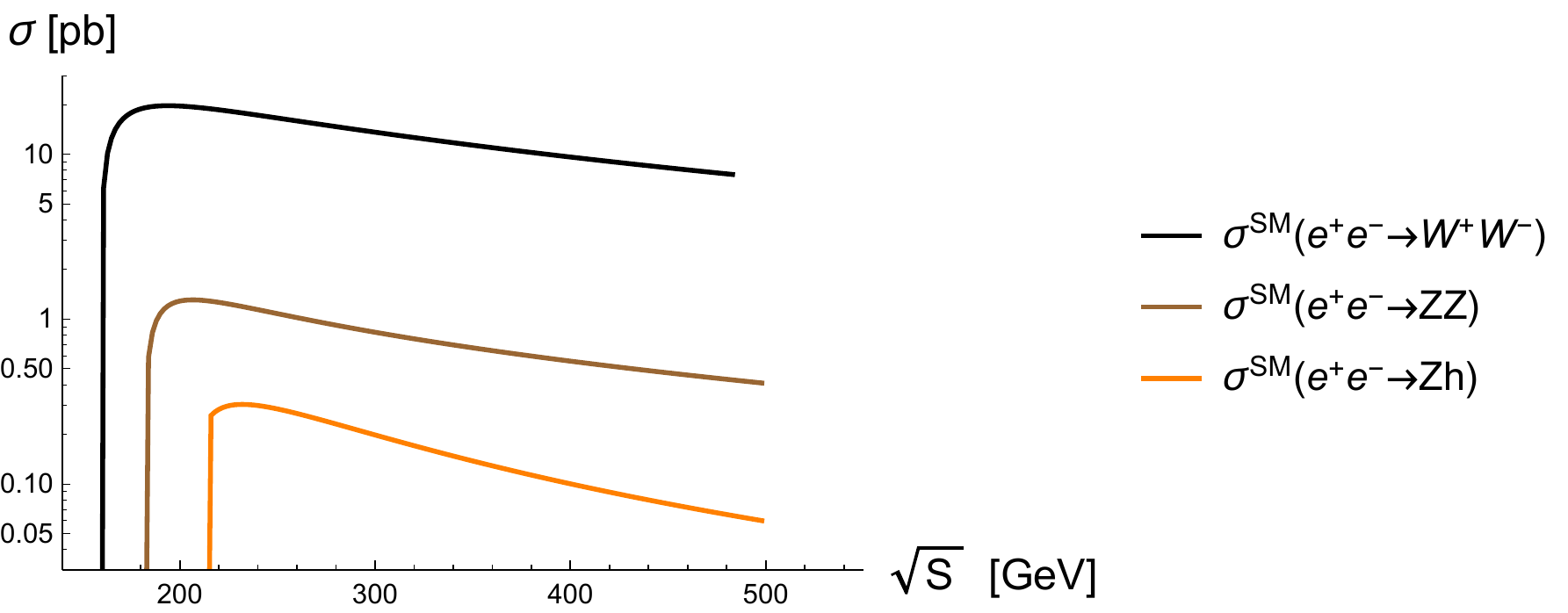}
\caption{The cross section for the SM background processes to our signal.}
\label{SM-background}
\end{center}
\end{figure}

With all our scalars lighter than 160 GeV, the final state of the $e^+e^- \to Z^* \to S_i S_j$ process would be missing transverse energy and $f \bar f$,
\bea 
&& e^+e^- \to Z^* \to S_1 S_j \to S_1 S_1 Z^* \to S_1 S_1 f \bar f,
 \\
&& e^+e^- \to Z^* \to S_i S_j \to S_1 Z^* S_1 Z^* \to S_1 S_1 f \bar f f \bar f,
\qquad (i,j=2,3,4).
\nonumber
\eea
The main SM background to such processes is through the channels
\be 
e^+ e^- \to Z Z \to f \bar f  \,\nu \bar\nu, \qquad
e^+ e^- \to W^+ W^- \to l^- \bar\nu  \,l^+ \nu, \qquad
e^+ e^- \to Z h \to f \bar f \, \slashed{E}_T\,.
\ee
Figure \ref{SM-background} shows the SM cross section for these processes in the energy range relevant to our analysis. Note that this background decreases with increasing energy, whereas the signal is asymptotically flat, as  shown in Figs. \ref{BP-A-fig}, \ref{BP-B-fig} \& \ref{BP-C-fig}.
Moreover, a simple calculation yields the cross-section for the aforementioned processes to be
\begin{footnotesize}
\bea
&& \sigma^{SM}(e^+ e^- \to W^+W^-) \times  \mbox{Br}(W^+ \to  l^+ \nu) \times \mbox{Br}(W^- \to  l^- \bar\nu) 
= 0.09 \;  \sigma^{SM}(e^+ e^- \to W^+W^-) \lessapprox 1.8  \; \mbox{[pb]},
\nonumber\\[2mm]
&& \sigma^{SM}(e^+ e^- \to Z Z) \times \mbox{Br}(Z \to f \bar f) \times \mbox{Br}(Z \to \nu \bar \nu) \times 2
= 0.28\;  \sigma^{SM}(e^+ e^- \to Z Z) \lessapprox 0.56 \; \mbox{[pb]},
\\[2mm]
&& \sigma^{SM}(e^+ e^- \to Z h) \times  \left[\mbox{Br}(Z \to f \bar f) \times \mbox{Br}(h \to inv.) 
+ \mbox{Br}(Z \to \nu \bar \nu) \times \mbox{Br}(h \to f \bar f) \right]
\nonumber\\[1mm]
&&
\hspace{10mm}
 = 0.28 \; \sigma^{SM}(e^+ e^- \to Z h) \lessapprox 0.112  \; \mbox{[pb]},
\nonumber
\eea
\end{footnotesize}
which should lead to a non-negligible significance of the signal over the background.

\section{Conclusion and outlook}
\label{conclusion}

We have studied distinctive signatures of the CP-violating I(2+1)HDM at a future $e^+e^-$ collider. The off-shell $Z$ boson in the process $e^+e^-\to Z^* \to S_iS_j$ leads to six possible final states involving pairs of dark (or inert) neutral states, $S_1S_{2,3,4}$, $S_2S_{3,4}$, $S_3S_4$, in the CP-violating case, in comparison to four possible final states, $H_1 A_{1,2}, H_2A_{1,2}$, in the CP-conserving case. We then have provided several BPs, for which we have shown production rates as large as a few picobarns at $\sqrt{s}$ energies accessible by future electron-positron colliders. The relative distance (in $\sqrt s$) of the production thresholds of these final states as well as their heights would serve the purpose of
separating typical dark scalar mass patterns, of which we benchmarked here three types, each corresponding to different DM dynamics compatible with relic density as well as both direct and indirect searches.  

Given the foreseen timescale for the construction and  exploitation stage of future $e^+ e^-$ colliders, we will therefore be able to probe the described I(2+1)HDM benchmark scenarios on  time scales of ten to twenty years from now. By that time, we expect an increased sensitivity of DM (in)direct detection experiments and more stringent constraints from a high energy and/or  luminosity LHC, so that, by combining information from all these sources, one may be in a position to eventually use extremely collimated and energetically precise electron-positron beams in order to perform a threshold scan able to accurately extract the six rest masses, $m_{S_i}+m_{S_j}$, in turn leading to a fit to the individual ones, $m_{S_i}$.

In fact, such a scope offered by future $e^+e^-$ colliders is complementary to the one that will be afforded by, e.g.,  a XENON$n$T upgrade. Furthermore, the former experiments are very useful for testing the I(2+1)HDM, as their measurements do not depend on the small Higgs-DM coupling, unlike the latter. Finally, 
it is worth noting that XENON$n$T (and other direct detection experiments) are sensitive only to the mass and  couplings of the DM particle and will provide no information about other unstable particles from the dark sector. Therefore, future $e^+e^-$ colliders will be essential to probe other inert particle masses than the DM candidate one, all of which will hardly be accessible at the LHC \cite{Preparation}, for the simple reason that, at an ILC \cite{Asner:2013psa} or FCC-$ee$ \cite{Gomez-Ceballos:2013zzn}, it is the highly controlled initial state that enable access to these while, at the LHC, this happens through final states always containing missing energy.

\section*{Acknowledgement}
The authors would like to thank Stephen F. King for numerous useful discussions.
SM acknowledges support from the STFC Consolidated grant ST/L000296/1 and is financed in part through the NExT Institute. SM and VK acknowledge the H2020-MSCA-RISE-2014 grant no. 645722 (NonMinimalHiggs).
DS is supported in part by the National Science Center, Poland, through the HARMONIA
project under contract UMO-2015/18/M/ST2/00518.
JH-S, DR and AC are supported by CONACYT (M\'exico), VIEP-BUAP and 
PRODEP-SEP (M\'exico) under the grant: ``Red Tem\'atica: F\'{\i}sica del Higgs y del Sabor". SM and AC thankfully acknowledge computer resources provided by Laboratorio Nacional de Supercomputo del Sureste de Mexico (LNS), a member of the CONACYT national laboratories, with project No. 201803024C.

\end{document}